\documentclass[twocolumn]{emulateapj}  
\usepackage{epstopdf}
\usepackage{ulem}
\usepackage{amssymb}
\usepackage{natbib}
\usepackage{times}
\usepackage{graphicx,bm,amssymb}
\usepackage{pstricks}

\usepackage{psfrag}
\usepackage[colorlinks=true,linkcolor=blue,citecolor=blue]{hyperref}
\newcommand{\be}{\begin{equation}}
\newcommand{\ee}{\end{equation}}
\newcommand{\bse}{\begin{subequations}}
\newcommand{\ese}{\end{subequations}}
\newcommand{\bary}{\begin{eqnarray}}
\newcommand{\eary}{\end{eqnarray}}

\shorttitle{}
\shortauthors{Fraija N.}

\begin{document}
\title{GRB 110731A: Early afterglow in stellar wind powered by a magnetized outflow}
\author{N. Fraija\altaffilmark{$\dagger$}}
\affil{Instituto de Astronom\'ia, Universidad Nacional Aut\'{o}noma de M\'{e}xico, Apdo. Postal 70-264, Cd. Universitaria, M\'{e}xico DF 04510}
\email{nifraija@astro.unam.mx}
\altaffiltext{$\dagger$}{Luc Binette-Fundaci\'on UNAM Fellow.}

\date{\today} 

\begin{abstract}
One of the most energetic gamma-ray burst GRB 110731A was observed from optical to GeV energy range.  Previous analysis on the prompt phase revealed similarities with the Large Area Telescope (LAT) bursts observed by Fermi: i) a delayed onset of the high-energy emission ($> 100$ MeV),  ii) a short-lasting bright peak at  later times and iii) a temporally extended component from this phase and  lasting hundreds of seconds. Additionally to the prompt phase, multiwavelength observations over different epochs showed that the spectral energy distribution was better fitted by a wind afterglow model.   We present a leptonic model  based on an early afterglow that evolves in a stellar wind of its progenitor.   We apply this model to interpret the temporally extended LAT emission and the brightest LAT peak exhibited by the prompt phase of GRB 110731A.  Additionally, using the same set of parameters, we describe the multiwavelength afterglow observations.  The origin of the temporally extended LAT, X-ray and optical flux is explained through synchrotron radiation from the forward shock and the brightest LAT peak is described evoking the  synchrotron self-Compton emission from the reverse shock. The bulk Lorentz factor required in this model ($\Gamma\simeq520$) lies in the range of values demanded for most LAT-detected gamma-ray bursts.  We show that the strength of the magnetic field in the reverse-shock region is $\sim$ 50 times stronger than in the forward-shock region. This result  suggests that for GRB 110731A,  the central engine is likely entrained with strong magnetic fields.  
\end{abstract}
\section{Introduction}
In recent years,  the detection of $\gamma$-rays and optical  polarization in gamma-ray bursts  (GRBs)  has supported the idea that jets could be magnetized \citep{2003Natur.423..415C, 2003astro.ph.10515B,2004MNRAS.350.1288R}.  The jet evolution with  magnetic  content has been explored in several contexts. In these models, an electromagnetic component  is introduced through the magnetization parameter ($\sigma$) and  defined by the ratio of Poynting flux (electromagnetic component) and matter energy (internal+kinetic component) \citep{2002A&A...387..714D,2003ApJ...596.1080V,2003ApJ...596.1104V,2000ApJ...537..810W,2002luml.conf..381B, 2002bjgr.conf..146L, 2004ASPC..312..357S, 2011ApJ...726...90Z, 2014ApJ...787..140F}.\\ 
Afterglow transition is one of the most interesting and least understood gamma-ray phases. During this phase, the relativistic ejecta interacts with the surrounding matter generating reverse and forward shocks.  A strong short-lived reverse shock (RS) propagates into the ejecta whereas the long-lasting forward shock (FS) leads to a continuous softening of the afterglow spectrum \citep{2007MNRAS.379..331P,2004MNRAS.353..647N}.  The dynamics of the RS in a wind and constant medium has been discussed by many authors \citep{2003ApJ...597..455K, 2003MNRAS.342.1131W, 1997ApJ...476..232M, 1999ApJ...517L.109S, 2012ApJ...751...33F,2012grb..confE..27F,2012ApJ...755..127S,2003ApJ...589L..69L}.  The RS has been invoked to explain the early $\gamma$-ray, optical and/or radio flares.    After the peak, no new electrons are injected and the material cools adiabatically, although if the central engine emits slowly moving material the RS could survive from hours to days \citep{2007MNRAS.381..732G, 2007ApJ...665L..93U}.   On the other hand, the origin of early flashes has been also discussed in the internal shock framework when  they are nearly two orders of magnitude weaker than those produced by RSs (for the same total energy) \citep{1997ApJ...476..232M,1999MNRAS.306L..39M,2000ApJ...532..286K}.   Early observations of GRB afterglows would offer to clarify the question whether the early emission takes place at internal or external shocks. \\ 
As has been pointed out in the literature,  emission regions and radiative processes of  high-energy (HE) photons with energies $\geq$ 100 MeV have been fully explored.  On hadronic models,  $\gamma$-ray components  have been explained through photo-hadronic interactions between HE hadrons accelerated in the jet  and internal synchrotron photons \citep{2009ApJ...705L.191A, 2000ApJ...537..255D},   inelastic proton-neutron collisions  \citep{2000ApJ...541L...5M}   and  interactions of HE neutrons and photons out of the jet \citep{2004A&A...418L...5D, 2004ApJ...604L..85A}.     On leptonic models,  $\gamma$-ray  fluxes have been explored with inverse Compton (IC), synchrotron self-Compton (SSC) and synchrotron  processes at different regions of the jet.  By considering electrons and photons at internal and external shocks,  IC emissions have been  discussed in detail in internal shocks \citep{1996ApJ...471L..91P,1998ApJ...494L.167P, 2000ApJ...544L..17P, 2007MNRAS.380...78G},  FSs \citep{1996ApJ...473..204S,1998ApJ...502L..13T,1997ApJ...489L..33W,1998ApJ...501..772P,1998ApJ...505..252W,1999ApJ...512..699C,2000ApJ...537..255D,2000ApJ...543...66P} and RSs \citep{2001ApJ...546L..33W,2001ApJ...556.1010W, 2006ApJ...643.1036P}.  GeV photons generated from SSC emission in FS   \citep{2001ApJ...548..787S,2001ApJ...546L..33W, 2001ApJ...559..110Z, 2012ApJ...755...12V, 2012ApJ...755..127S} and RS \citep{2003ApJ...598L..11G,2001ApJ...546L..33W,2001ApJ...556.1010W, 2012ApJ...751...33F} have been investigated  separately  and  only in few cases  synchrotron radiation  has been  examined \citep{2011ApJ...730....1L,2011ApJ...733...22H,2010ApJ...718L..63P, 2009MNRAS.400L..75K, 2010MNRAS.409..226K,2010MNRAS.403..926G}.    Additionally, a particular lepto-hadronic model was developed to explain the HE emissions of GRB090510 \citep{2010ApJ...724L.109R}.\\ 
The bright and long  GRB 110731A was detected by Fermi and Swift observatories, and by the MOA and GROND optical telescopes.  The analysis of the prompt phase  revealed a brightest peak in the LAT light curve  starting at $\sim$ 5.5 s \citep{2013ApJ...763...71A} and a temporally extended LAT component described with a power law.  In addition, temporal and spectral analysis in different wavelengths and epochs  (just after the trigger time and extending for more than 800 s) favored a wind afterglow  model.   Recently,  assuming that the long-lasting LAT component could be described as synchrotron radiation by relativistic electrons accelerated through FSs and requiring that the magnetic equipartition parameter varies as a function of time, $\epsilon_B\propto t^{-\alpha_t}$ with  $0.5\leq\alpha_t \leq 0.4$, \cite{2013arXiv1305.3689L} interpreted that these GeV photons were likely produced in a region of strong $\epsilon_B$. They argued that the magnetization that permeates the blast wave of GRB 110731A can be described as partial decay of the micro-turbulence \citep{2003MNRAS.339..881R, 2006ApJ...653..454P} as observed in particle-in-cell (PIC) simulations \citep{2008ApJ...682L...5S, 2009ApJ...695L.189M,  2011ApJ...739L..42H, 2013ApJ...771...54S, 2011ApJ...726...75S}.\\
In this paper, we develop a leptonic model based on early afterglow with variable density (stellar wind, s=2)  to describe the temporally extended emission  and  the brightest peak  present in the LAT light curve of GRB 110731A. The paper is arranged as follows. In Section 2 we show a leptonic model  based on external shocks (forward and reverse) that evolves adiabatically in a stellar wind.  In section 3 we apply this model to GRB 110731A as a particular case and in Section 4 we give a brief summary.  
\section{External Shock Model}
As the blast wave extends out into the stellar dense wind of the progenitor, it starts to be decelerated leading to forward and reverse shocks. The evolution of the afterglow will mainly depend on its mass and in some cases, the emission processes of internal and reverse shocks which could be simultaneously present in the light curve.  In the following subsections we will develop the dynamics of the external shocks in stellar winds when Fermi-accelerated electrons are cooled down by synchrotron and Compton scattering emission at forward and reverse shocks. In addition, we will consider the RS in the thick- and thin-shell case. We hereafter use primes (unprimes) to define the quantities in a comoving (observer) frame and  c=$\hbar$=1 in natural units. The subscripts f and r refer throughout this paper to the forward and reverse shocks, respectively.
\subsection{Forward Shocks}
Afterglow hydrodynamics involves a relativistic blast wave expanding into the medium with density
\be\label{rho}
\rho=A\,r^{-2} \hspace{0.3cm} {\rm with}  \hspace{0.3cm}  A=\frac{\dot{M}_w}{4\pi V_w}\,,
\ee
where $\dot{M}_w$ is the mass loss rate and $V_w$ is the wind velocity.  For an ultra relativistic and adiabatic blast wave, the radius shock ($r$) spreading into this density can be written in the form
\be\label{rad}
r=\frac{3\xi}{2\pi^{1/2}} (1+z)^{-1/2}\,E^{1/2}\,t^{1/2}\,A^{-1/2}\,.
\ee
Here the total energy ($E$) of the shock is constant and given by $E=8\pi/9\,A\,\Gamma^2\,r$, $\Gamma$ is the bulk Lorentz factor, $\xi$ is a constant parameter \citep{1998ApJ...493L..31P}, $z$ is the redshift and $t$ is the time in the observer's frame  \citep{1998MNRAS.298...87D, 1997ApJ...489L..37S, 1998ApJ...493L..31P} which is given by
\be\label{tim}
t=(1+z)\,\frac{r}{4\,\xi^2\,\Gamma^2}\,.
\ee
From eqs. (\ref{rho}), (\ref{rad}) and (\ref{tim}), we get the scale of deceleration time 
\be\label{t_dec}
t_{dec}=\frac{9}{64\pi\,\xi^2}(1+z)\,E\,A^{-1}\,\Gamma^{-4}.
\ee
\paragraph{Synchrotron emission}. 
Considering that electrons are accelerated to a power-law distribution $N(\gamma_e) d\gamma_e\propto \gamma_e^{-p} d\gamma_e$  with the electron spectral index p$>$ 2,  and the energy density ($U$) is equipartitioned to accelerate electrons ($U_e= \epsilon_{e,f}\,U=m_e \int\gamma_e N(\gamma_e)d\gamma_e$) and to amplify the magnetic field   $U_{B,f} = \epsilon_{B,f}\,U$  (with $U_{B,f}=B'^2_f/8\pi$),    the minimum electron Lorentz factor and the magnetic field can be written as
\be\label{gamma_m}
\gamma_{e,m,f}=\frac{(p-2)\,m_p}{(p-1)\,m_e} \epsilon_{e,f}\,\Gamma\,,
\ee
and
\be
B'_f\simeq \frac{8\sqrt2\,\pi}{3\xi}(1+z)^{1/2} \epsilon_{B,f}^{1/2}\,\Gamma\,E^{-1/2}\,t^{-1/2}\,A\,,
\ee
respectively.  Here, $\epsilon_{e,f}$ ($\epsilon_{B,f}$) is the electron (magnetic) equipartition parameter and  $m_e$ ($m_p$) is the electron (proton) mass.  When  the expanding relativistic ejecta encounter the stellar wind, it starts to be decelerated, then electrons are firstly heated and after cooled down  by synchrotron emission. Comparing both time scales, the deceleration time (eq. \ref{t_dec})     and the characteristic cooling time for synchrotron radiation  {\small $t_{e,syn}\simeq 3m_e/(16\sigma_T)\,(1+x_f)^{-1}\,(1+z)\,\epsilon^{-1}_{B,f}\,\rho^{-1}\,\Gamma^{-3}\,\gamma_e^{-1}$}, the characteristic electron Lorentz factor can be written in the form
{\small
\be\label{gamma_c}
\gamma_{e,c,f}=\frac{3m_e\xi^4}{\sigma_T}\,(1+x_f)^{-1}\,(1+z)^{-1}\,\epsilon^{-1}_{B,f}\,\Gamma\,A^{-1}\,t\,.
\ee
}
Here $\sigma_T$ is the Thomson cross section and the term $(1+x_f)$ is introduced because a once-scattered synchrotron photon generally has energy larger than the electron mass in the rest frame of the second-scattering electrons. It is given by \citep{2001ApJ...548..787S}
\begin{equation}\label{xf}
1+x_f = \left\{ \begin{array} {ll} 
1+\frac{\eta \epsilon_{e,f}}{\epsilon_{B,f}},                            &      \mathrm{if \quad} \frac{\eta \epsilon_{e,f}}{\epsilon_{B,f}} \ll 1, \\ 
1+\left(\frac{\eta \epsilon_{e,f}}{\epsilon_{B,f}}\right)^{1/2},   &      \mathrm{if \quad} \frac{\eta \epsilon_{e,f}}{\epsilon_{B,f}} \gg 1\,,
\end{array} \right.
\end{equation}
where 
\begin{equation}\label{eta}
\eta = \left\{ \begin{array} {ll} 
\left(\frac{\gamma_{\rm e,c,f}}{\gamma_{\rm e,m,f}}\right)^{2-p}\,, &      \mathrm{for\,\, slow\,\, cooling}\,, \\
1\,,                                                                          &      \mathrm{for\,\, fast\,\, cooling}\,.
\end{array} \right.
\end{equation}
The maximum electron Lorentz factor can be calculated comparing  the acceleration $t_{acc}\simeq \frac{2\pi\,m_e}{q_e}(1+z)\,\Gamma^{-1}\,{B'}^{-1}_f \gamma_e$ and cooling ($t_{e,syn}$) time scales. Hence, the maximum electron Lorentz factor is
\bary\label{gamma_max}
\gamma_{e,max,f}&\simeq& \sqrt{\frac{9\sqrt2\,q_e}{16\,\pi \sigma_T}}\,\xi^{1/2}\,(1+z)^{-1/4}\epsilon_{B,f}^{-1/4}\Gamma^{-1/2}E^{1/4}\cr
&&\hspace{3.8cm}\times\,A^{-1/2}\,t^{1/4}\,,
\eary
where $q_e$ is the elementary charge.  From eqs. (\ref{gamma_m}), (\ref{gamma_c}), (\ref{gamma_max}) and (\ref{t_dec}), the synchrotron spectral breaks  computed through the synchrotron emission $E_{i,f}=\frac{q_e}{m_e}\,(1+z)^{-1}\,\Gamma\,B'\gamma^2_{e,i,f}$ for i=m,c and max can be written as
{\small
\bary\label{synforw}
E^{syn}_{\rm \gamma,a,f} &\simeq& \frac{2^{16/5}\pi^{21/10}\,3^{1/5} q_e^{8/5}\,(p+2)^{3/5}\,(p-1)^{8/5}\,\xi^{-6/5}}{5^{3/5}\,\Gamma(5/6)^{3/5}\,(3p+2)^{3/5}(p-2)\,m_p^{8/5}}\cr
&&\hspace{1.3cm}\times\,(1+z)^{-2/5}\,\epsilon_{e,f}^{-1}\,\epsilon_{B,f}^{1/5}\, E^{-2/5}\,A^{6/5} t^{-3/5}\cr
E^{syn}_{\rm \gamma,m,f} &\simeq& \frac{3\sqrt2q_e\,m_p^2\,(p-2)^2}{8m_e^3\,\xi^3\,(p-1)^2} (1+z)^{1/2}\,\epsilon_{e,f}^2\,\epsilon_{B,f}^{1/2}\,E^{1/2}\,  t^{-3/2}\cr
E^{syn}_{\rm \gamma,c,f}  &\simeq& \frac{27\sqrt{2\pi}\,q_em_e\,\xi^5}{8\sigma^2_T}  (1+z)^{-3/2}\,(1+x_f)^{-2}\,\epsilon_{B,f}^{-3/2}\,\cr
&& \hspace{4.4cm}\times\, A^{-2}\,E^{1/2}\, t^{1/2}\, \cr
E^{syn}_{\rm \gamma,max,f}  &\simeq& \frac{3\sqrt{3}\,q_e^2\,\xi^{-1/2}}{2\sqrt2\,\pi^{1/4} m_e\sigma_T}\,(1+z)^{-3/4}\,E^{1/4}\,A^{-1/4}\,t^{-1/4}\,.
\eary
}
The synchrotron self-absorption energy $E^{syn}_{\rm \gamma,a,f}$ was calculated through the absorption coefficient  \citep{1986rpa..book.....R}
{\small
\bary
\alpha_{\rm \epsilon'_a} &\simeq& \frac{2^{22/3}\pi^{25/6}\,q_e^{8/3}\,(p+2)\,(p-1)^{8/3}\,\xi^{-8/3}}{15\,\Gamma(5/6)\,(3p+2)(p-2)^{5/3}\,m_p^{8/3}}\,(1+z)^{4/3}\cr
&&\hspace{1.2cm}\times\,\epsilon_{e,f}^{-5/3}\,\epsilon_{B,f}^{1/3}\, E^{-4/3}\,A^{8/3} t^{-4/3} {E'^{syn}_{\rm \gamma,a,f}}^{-5/3}\,, 
\eary
}
and the condition $\alpha_{\rm \epsilon'_a} r/\Gamma=1$ \citep{1999ApJ...523..177W,1999ApJ...527..236G}.  The maximum flux  $F^{syn}_{\rm \gamma,max,f}=N_e P_{\nu,max}/4\pi D^2$ with peak spectral power  $P_{\nu,max}\simeq  \sigma_T(m_e/3q_e)\,(1+z)^{-1}\,\Gamma\,B'_f$  can be written as 
\be\label{Fsyn}
F^{syn}_{\rm \gamma,max,f}\simeq \frac{\sqrt{2\pi} m_e\sigma_T}{24\, q_e\,m_p\,\xi}  (1+z)^{3/2}\,\epsilon_{B,f}^{1/2} \,A\,D^{-2}\,E^{1/2}\,t^{-1/2}\,, 
\ee
where $D$ is the luminosity distance from the source.  Following \citet{1998ApJ...497L..17S},  the observed synchrotron spectrum can be derived in the fast- and slow-cooling regime as 
{\small
\begin{eqnarray}
\label{fcsyn}
&&F^{syn}_\nu (E^{syn}_\gamma)=
F^{syn}_{\rm \gamma,max,f}\hspace{6cm}\cr
&&\hspace{0.2cm}\times\cases{ 
\left( \frac{E^{syn}_\gamma}{E^{syn}_{\rm \gamma,c,f}}\right)^{-\frac12}, \hspace{1.3cm}  E^{syn}_{\rm \gamma,c,f}<E^{syn}_\gamma<E^{syn}_{\rm \gamma,m,f}, \cr
\left(\frac{E^{syn}_{\rm \gamma,m,f}}{E_{\rm \gamma,c,f}}\right)^{-\frac12}  \left(\frac{E^{syn}_\gamma}{E^{syn}_{\rm \gamma,m,f}}\right)^{-\frac{p}{2}}\,,E^{syn}_{\rm \gamma,m,f}<E^{syn}_\gamma<E^{syn}_{\rm \gamma,max,f}. \cr
}
\end{eqnarray}
}
\noindent and
{\small
\begin{eqnarray}
\label{scsyn}
&&F^{syn}_\nu(E^{syn}_\gamma)=
F^{syn}_{\rm \gamma,max,f}\hspace{6cm}\cr
&&\times \cases{ 
\left(\frac{E^{syn}_\gamma}{E^{syn}_{\rm \gamma,m,f}}\right)^{-\frac{p-1}{2}},\hspace{1cm}  E^{syn}_{\rm \gamma,m,f}<E^{syn}_\gamma<E^{syn}_{\rm \gamma,c,f}, \cr
\left(\frac{E^{syn}_{\rm \gamma,c,f}}{E^{syn}_{\rm \gamma,m,f}}\right)^{-\frac{p-1}{2}} \left(\frac{E^{syn}_\gamma}{E^{syn}_{\rm \gamma,c,f}}\right)^{-\frac{p}{2}}\,,  E^{syn}_{\rm \gamma,c,f}<E^{syn}_\gamma<E^{syn}_{\rm \gamma,max,f}\,, \cr
}
\end{eqnarray}
}
respectively. Here $F^{syn}_{\rm \gamma,max,f}$ and $E^{syn}_{\rm \gamma,i,f}$ are given by eqs. (\ref{Fsyn}) and (\ref{synforw}), respectively.  The transition time ($t^{syn}_0$) from fast- to slow-cooling spectrum is 
{\small
\bary\label{t0}
t^{syn}_0=\frac{\sigma_T\,m_p(p-2)}{3\pi^{1/4}\,m_e^2(p-1)\xi^4}\,(1+z)\epsilon_{e,f}\,\epsilon_{B,f}\,A\,.
\eary
}
Using the synchrotron spectral breaks  radiated (eq. \ref{synforw}) and synchrotron spectrum in the fast (eq \ref{fcsyn}) and slow (eq \ref{scsyn}) cooling regime, one can obtain the light curves (LCs) as a function of energy ($E^{syn}_\gamma$). We get the flux for fast-cooling regime
{\small
\begin{eqnarray}
\label{fcsyn_t}
F^{syn}_\nu= \cases{ 
A^{syn}_{fl}\,t^{-\frac{1}{4}}\, \left(E^{syn}_\gamma\right)^{-\frac12},\,\,\,\, E^{syn}_{\rm \gamma,c,f}<E^{syn}_\gamma<E^{syn}_{\rm \gamma,m,f}, \cr
A^{syn}_{fh}\,t^{-\frac{3p-2}{4}}\,\left(E^{syn}_\gamma\right)^{-\frac{p}{2}},\,\,\,\,E^{syn}_{\rm \gamma,m,f}<E^{syn}_\gamma<E^{syn}_{\rm \gamma,max,f}. \cr
}
\end{eqnarray}
}
where $A^{syn}_{fl}$ and $A^{syn}_{fh}$ are
{\small
\bary\label{A1}
A^{syn}_{fl}&\simeq&\left(\frac{(2\pi)^{3/4}\,27^{1/2}\,m_e^{3/2}\,\xi^{3/2}}{248^{1/2}\,q_e^{1/2} m_p}\right)\, (1+z)^{\frac{3}{4}}\,(1+x_f)^{-1}\,\epsilon_{e,f}^{-\frac34}\cr
&&\hspace{4.5cm}\times\,\epsilon_{B,f}^{\frac{1}{2}}\, E^{\frac{3}{4}}\,D^{-2}
\eary
}
and
{\small
\bary\label{A2}
A^{syn}_{fh}&\simeq&\left(\frac{2^{1/2}\pi^{3/4}m_e^3\xi^3(p-1)}{4q_em_p^2(p-2)}\right)\left(\frac{3\sqrt2 q_em_p^2(p-2)^2}{8m_e^3\xi^3(p-1)^2}\right)^\frac{p}{2} \cr
&&\hspace{0.3cm}\times\,(1+x_f)^{-1} \,(1+z)^{\frac{p+2}{4}}    \epsilon_{e,f}^{p-1}\,\epsilon_{B,f}^{\frac{p-2}{4}}\, E^{\frac{p+2}{4}}\,D^{-2}\,,
\eary
}
\noindent respectively.  For  slow-cooling regime, we get 
{\small
\begin{eqnarray}
\label{scsyn_t}
F^{syn}_\nu=\cases{
A^{syn}_{sl}t^{-\frac{3p-1}{4}}\left(E^{syn}_\gamma\right)^{-\frac{p-1}{2}},E^{syn}_{\rm \gamma,m,f}<E^{syn}_\gamma<E^{syn}_{\rm \gamma,c,f},\cr
A^{syn}_{fh}\,t^{-\frac{3p-2}{4}}\,\left(E^{syn}_\gamma\right)^{-\frac{p}{2}},\,E^{syn}_{\rm \gamma,c,f}<E^{syn}_\gamma<E^{syn}_{\rm \gamma,max,f}. \cr
}
\end{eqnarray}
}
with $A^{syn}_{sl}$ given by
{\small
\bary\label{A3}
A^{syn}_{sl}&=&\left(\frac{\sqrt{2\pi}m_e\sigma_T}{24q_em_p \xi}\right)\left(\frac{3\sqrt2 q_em_p^2(p-2)^2}{8m_e^3\xi^3(p-1)^2}\right)^\frac{p-1}{2} (1+z)^{\frac{p+5}{4}} \cr
&&\hspace{3.0cm}\times\,\epsilon_{e,f}^{p-1}\,\epsilon_{B,f}^{\frac{p+1}{4}}\,A\, E^{\frac{p+1}{4}}\,D^{-2}\,. 
\eary
}
\paragraph{SSC emission}.
Fermi-accelerated electrons in the FS may scatter synchrotron photons up to higher energies $E^{ssc}_{\gamma,i}\simeq 2 \gamma^2_{e,i} E^{syn}_{\gamma,i}$. From the synchrotron spectral breaks (eqs. \ref{synforw}) and break electron Lorentz factors (eqs. \ref{gamma_m}, \ref{gamma_c} and \ref{gamma_max}), the spectral breaks in the Compton regime are
{\small
\bary\label{icforw}
E^{ssc}_{\rm \gamma,a,f} &\simeq& \frac{2^{1/5}\pi^{8/5}\,3^{6/5} q_e^{8/5}\,m_p^{2/5} (p+2)^{3/5}\,(p-2)\,\xi^{-11/5}}{5^{3/5}\,\Gamma(5/6)^{3/5}\,m_e^2(3p+2)^{3/5}(p-1)^{2/5}}\cr
&&\hspace{1.3cm}\times\,(1+z)^{1/10}\,\epsilon_{e,f}\,\epsilon_{B,f}^{1/5}\, E^{1/10}\,A^{7/10} t^{-11/10}\cr
E^{ssc}_{\rm \gamma,m,f} &\simeq&\frac{9\sqrt2\,q_e\,m_p^4(p-2)^4}{64\,\pi^{1/2}m^5_e\,\xi^4(p-1)^4} (1+z)\,\epsilon_{e,f}^4\,\epsilon_{B,f}^{1/2}\,A^{-1/2}\,E\,  t^{-2}\cr
E^{ssc}_{\rm \gamma,c,f}  &\simeq&\frac{729\sqrt2\,q_e\,m_e^3}{64\,\sigma_T^4}\,\xi^{12}(1+z)^{-3}\,(1+x_f)^{-4}\,\epsilon_{B,f}^{-7/2}\cr
&& \hspace{5cm}\times\, A^{-9/2}\,E\, t^{2}\,\cr 
E^{ssc}_{\rm \gamma,max,f}  &\simeq&\frac{27\sqrt3\,q^3_e\,\xi^{1/2}}{32  \pi^{5/4}\sigma_T^2\,m_e}\,(1+z)^{-5/4}\,\epsilon_{B,f}^{-1/2}\,\Gamma^{-1}\,E^{3/4}\cr
&& \hspace{4.5cm}\times\,A^{-5/4}\, t^{1/4}\,.
\eary
}
Here we have calculated the self-absorption energy in the Compton regime through $E^{ssc}_{\rm \gamma,a,f} \simeq \gamma_m^2 E^{syn}_{\rm \gamma,a,f}$ \citep{2001ApJ...548..787S}.  From eqs. (\ref{rho}),  (\ref{rad}) and (\ref{Fsyn}), we get  
\bary
F^{ssc}_{\rm \gamma,max,f}&\simeq&\,(\sigma_T/m_p)\,r\,\rho\,F^{syn}_{\rm \gamma,max,f}\cr
&\simeq&\frac{\sqrt2 m_e\sigma^2_T}{36\,q_e\,m^2_p\,\xi^2} (1+z)^2\,\epsilon_{B,f}^{1/2} \,A^{5/2}\,D^{-2}\,t^{-1}\,.
\eary
In the Klein-Nishina (KN) regime, the emissivity of IC radiation per electron is independent of the electron energy and reduced in comparison with the classical regime. The observed KN break energy is
{\small
\bary
E^{KN}_{\rm \gamma,f}&\simeq& \frac{\Gamma}{1+z}\gamma_{e,c,f}\,m_e\cr
&\simeq& \frac{3m^2_e\xi^4}{\sigma_T}\,(1+x_f)^{-1}\,(1+z)^{-2}\,\epsilon^{-1}_{B,f}\,\Gamma^2\,A^{-1}\,t\,,
\eary
}
where we have used eqs. (\ref{t_dec}) and (\ref{gamma_c}).  From the synchrotron spectra (eqs. \ref{fcsyn} and \ref{scsyn}), we get  the SSC spectra for  the fast-cooling regime 
\vspace{0.45cm}
{\small
\begin{eqnarray}
\label{fcic}
&&E^{ssc}_\gamma F^{ssc}_{\rm \nu,f} (E^{ssc}_\gamma)=
x_f\,(E_\gamma F_\gamma)^{syn}_{\rm max,f}\hspace{4cm}\cr
&&\hspace{1cm}\times\cases{ 
\left( \frac{E^{ssc}_\gamma}{E^{ssc}_{\rm \gamma,c,f}}\right)^{\frac{1}{2}}\hspace{0.5cm} , \hspace{0.2cm}  E^{ssc}_{\rm \gamma,c,f}<E^{ssc}_\gamma<E^{ssc}_{\rm \gamma,m,f}, \cr
  \left(\frac{E^{ssc}_\gamma}{E^{ssc}_{\rm \gamma,m,f}}\right)^{\frac{2-p}{2}}\,,E^{ssc}_{\rm \gamma,m,f}<E^{ssc}_\gamma < E^{ssc}_{\rm \gamma,max,f}\,, \cr
}
\end{eqnarray}
}
\noindent where $(E_\gamma F_\gamma)^{syn}_{\rm max,f}=  E^{syn}_{\rm \gamma,m,f}  F^{syn}_\nu(E^{syn}_{\rm \gamma,m,f})$ and for the slow-cooling regime
{\small
\begin{eqnarray}
\label{scic}
&&E^{ssc}_\gamma F^{ssc}_{\rm \nu,f}(E^{ssc}_\gamma) =
x_f\,(E_\gamma F_\gamma)^{syn}_{\rm max,f}\hspace{4cm}\cr
&&\hspace{1cm}\times \cases{ 
\left(\frac{E^{ssc}_\gamma}{E^{ssc}_{\rm \gamma,m,f}}\right)^{\frac{3-p}{2}},\hspace{0.2cm}  E^{ssc}_{\rm \gamma,m,f}<E^{ssc}_\gamma<E^{ssc}_{\rm \gamma,c,f}, \cr
\left(\frac{E^{ssc}_\gamma}{E^{ssc}_{\rm \gamma,c,f}}\right)^{\frac{2-p}{2}}\,, E^{ssc}_{\rm \gamma,c,f}<E^{ssc}_\gamma  < E^{ssc}_{\rm \gamma,max,f}\,,\cr
}
\end{eqnarray}
}
with $x_f$ given by eqs. (\ref{xf}) and (\ref{eta}), and $(E_\gamma F_\gamma)^{syn}_{\rm max,f}=  E^{syn}_{\rm \gamma,c,f}  F^{syn}_\nu (E^{syn}_{\rm \gamma,c,f})$ \citep{2001ApJ...548..787S,2003ApJ...598L..11G}.  From the break photon energies (eq. \ref{icforw}) and synchrotron spectrum in the fast (eq. \ref{fcic}) and slow (eq. \ref{scic}) cooling regime, one can obtain the LCs for the fast- cooling
{\small
\begin{eqnarray}
\label{fcic_t}
F^{ssc}_\nu\sim\cases{ 
t^0\, \left(E^{ssc}_\gamma\right)^{-\frac12},\,\,E^{ssc}_{\rm \gamma,c,f}<E^{ssc}_\gamma<E^{ssc}_{\rm \gamma,m,f}, \cr
t^{-p+1}\,\left(E^{ssc}_\gamma\right)^{-\frac{p}{2}},\,\,E^{ssc}_{\rm \gamma,m,f}<E^{ssc}_\gamma < E^{ssc}_{\rm \gamma,max,f}. \cr
}
\end{eqnarray}
}
\noindent and the slow-cooling regime
{\small
\begin{eqnarray}
\label{scic_t}
F^{ssc}_\nu\sim\cases{ 
t^{-p}\, \left(E^{ssc}_\gamma\right)^{-\frac{p-1}{2}},E^{ssc}_{\rm \gamma,m,f}<E^{ssc}_\gamma<E^{ssc}_{\rm \gamma,c,f},\cr
t^{-p+1}\,\left(E^{ssc}_\gamma\right)^{-\frac{p}{2}},\, E^{ssc}_{\rm \gamma,c,f}<E^{ssc}_\gamma  < E^{ssc}_{\rm \gamma,max,f}\,,\cr
}
\end{eqnarray}
}
as a function of energy ($E^{ssc}_\gamma$).
\subsection{Reverse Shocks}
For the RS, a simple analytic solution can be derived taking two limiting cases, thick- and thin-shell case,  \citep{1995ApJ...455L.143S} by using a critical Lorentz factor ($\Gamma_c$) which is defined by 
\bary
\Gamma_c&=&\sqrt{\frac{3}{8\,\pi^{1/2}\,\xi}}\,(1+z)^{1/4}\,E^{1/4}\,A^{-1/4}\,T^{-1/4}_{90}\,,
\eary
where  $T_{90}$ is the duration of the GRB.  For $\Gamma>\Gamma_c$ (thick shell) the shell is significantly decelerated by the RS\footnote{Although bulk Lorenz factors (FS and RS) can be different at the shocked region, we have considered them to be similar.}\, otherwise, $\Gamma<\Gamma_c$ (thin shell), the RS cannot decelerate the shell effectively.    Irrespective of the evolution of RS,  the synchrotron  spectral evolution between RS and FS is related by \citep{2003ApJ...595..950Z,2007ApJ...655..973K,2005MNRAS.364L..42F,2004MNRAS.351L..78F,2007MNRAS.378.1043J,2005ApJ...633.1027S} 
\bary\label{conec}
E^{syn}_{\rm \gamma, m,r}&\sim&\,\mathcal{R}^2_e\,\mathcal{R}^{-1/2}_B\,\mathcal{R}^{-2}_M\,E^{syn}_{\rm \gamma,m,f}\cr
E^{syn}_{\rm \gamma,c,r}&\sim&\,\mathcal{R}^{3/2}_B\,\mathcal{R}^{-2}_x\,E^{syn}_{\rm \gamma,c,f}\cr
F^{syn}_{\rm \gamma,max,r}&\sim&\,\mathcal{R}^{-1/2}_B\,\mathcal{R}_M\,F^{syn}_{\gamma,max,f}
\eary
where
{\small
\be\label{param}
\mathcal{R}_B=\frac{\epsilon_{B,f}}{\epsilon_{B,r}},\hspace{0.1cm} \mathcal{R}_e=\frac{\epsilon_{e,r}}{\epsilon_{e,f}},\hspace{0.1cm}  \mathcal{R}_x=\frac{1+x_f}{1+x_r+x_r^2}\hspace{0.1cm} {\rm and}\hspace{0.1cm} \mathcal{R}_M=\frac{\Gamma^2_{d}}{\Gamma}\,,
\ee
}
where  $\Gamma_{d}$ is the bulk Lorentz factor at the shock crossing time. The previous relations tell us that including the re-scaling there is a unified description between forward and reverse shocks, and the distinction between forward and reverse magnetic fields considers that in some central engine models  \citep{1992Natur.357..472U, 1997ApJ...482L..29M,2000ApJ...537..810W} the fireball could be endowed with '"primordial" magnetic fields.  Also as the cooling Lorentz factor must be corrected, then  $\mathcal{R}_x$ is introduced as a correction factor for the SSC cooling, where $x_r$ is obtained by \citep{2007ApJ...655..973K}
\begin{equation}
x_r = \left\{ \begin{array} {ll} 
\frac{\eta \epsilon_{e,r}}{\epsilon_{B,r}}, & \mathrm{if \quad}
\frac{\eta \epsilon_{e,r}}{\epsilon_{B,r}} \ll 1, \\ 
\left(\frac{\eta \epsilon_{e,r}}{\epsilon_{B,r}}\right)^{1/3}, & \mathrm{if \quad}
\frac{\eta \epsilon_{e,r}}{\epsilon_{B,r}} \gg 1. 
\end{array} \right.
\end{equation}
Here $\eta=(\gamma_{\rm e,c,r}/\gamma_{\rm e,m,r})^{2-p}$ is given for slow-cooling and $\eta=1$ for fast-cooling regime. 
\subsubsection{Thick-shell case}
In this case,  the RS becomes relativistic during its propagation and the ejecta is significantly decelerated.  The bulk Lorentz factor at the shock crossing time  $t_d\simeq T_{90}$ is given by $\Gamma_{d}\sim \Gamma_c$.  Eventually,  the shock crossing time could be shorter than $T_{90}$ depending on the degree of magnetization of ejecta, defined as the ratio of Poynting flux to matter energy flux  $\sigma =L_{pf}/L_{kn}\sim \epsilon_{B,r}$  \citep{2004A&A...424..477F,2005ApJ...628..315Z,2007ApJ...655..973K}.  In particular, numerical analysis performed by \cite{2004A&A...424..477F} revealed that for the value of the magnetization parameter $\sigma \simeq 1$, the shock crossing time becomes $t_d\simeq T_{90}/6$.\\
\paragraph{Synchrotron emission}. Assuming that electrons are accelerated in the RS to a power-law distribution and the energy density is equipartitioned  between electrons and magnetic field, then the minimum electron Lorentz factor  and magnetic field are 
{\small
\bary\label{gamma_mr}
\gamma_{\rm e,m,r}&=&\epsilon_{e,r}\biggl(\frac {p-2} {p-1}\biggr) \frac {m_p} {m_e}\frac{\Gamma}{\Gamma_{d}}\cr
&=&\frac{\sqrt8\pi^{1/4}m_p(p-2)\,\xi^{1/2}} {\sqrt3m_e(p-1)}\, (1+z)^{-1/4}\,\epsilon_{e,r}\,  \Gamma\,A^{1/4}\,E^{-1/4}\cr
&& \hspace{5.8cm}\times\, T^{1/4}_{90}\,,
\eary
}
and
\be
B'_r\simeq \frac{8\sqrt2\,\pi}{3\xi}(1+z)^{1/2} \epsilon_{B,r}^{1/2}\,\Gamma\,E^{-1/2}\,T_{90}^{-1/2}\,A\,.
\ee
From the characteristic cooling time of synchrotron radiation and dynamical time scale, the characteristic electron Lorentz factor can be written as
{\small
\be\label{gamma_cr}
\gamma_{e,c,r}=\frac{27m_e\,\xi^2}{64\pi \sigma_T}\,(1+x_r+x^2_r)^{-1}\,\epsilon^{-1}_{B,r}\,E\,\Gamma^{-3}\,A^{-2}.
\ee
}
By considering $\gamma_{e,a}\simeq\gamma_{e,m}$ \citep{2001ApJ...548..787S} and from eq. (\ref{param}), we re-scale the synchrotron self-absorption energy between FS and RS as {\small $E^{syn}_{\rm \gamma, a,r}\sim\,\mathcal{R}^2_e\,\mathcal{R}^{-1/5}_B\,\mathcal{R}^{-2}_M\,E^{syn}_{\rm \gamma,a,f}$}. Additionally,  from eqs. (\ref{synforw}), (\ref{param}) and (\ref{conec}), we get the synchrotron spectral breaks  
{\small
\bary\label{synrev}
E^{syn}_{\rm \gamma,a,r} &\simeq& \frac{2^{46/5}\pi^{31/10} q_e^{8/5}\,(p+2)^{3/5}\,(p-1)^{8/5}\,\xi^{4/5}}{5^{3/5}\,3^{9/5}\,\Gamma(5/6)^{3/5}\,(3p+2)^{3/5}(p-2)\,m_p^{8/5}}\cr
&&\hspace{1.3cm}\times\,(1+z)^{-7/5}\,\epsilon_{e,r}^{-1}\,\epsilon_{B,r}^{1/5}\, E^{-7/5}\,A^{11/5}\,\Gamma^2 T^{2/5}_{90}\cr
E^{syn}_{\rm \gamma, m,r}&\simeq& \frac{8\pi\sqrt2\,q_e\,m_p^2\,(p-2)^2}{3\,\,m_e^3\,\xi(p-1)^2}\,(1+z)^{-1/2}\,\epsilon_{e,r}^{2}\,\epsilon_{B,r}^{1/2}\,\Gamma^{2}\,E^{-1/2}\cr
&&\hspace{5.8cm}\times\,A\,T_{90}^{-1/2} \cr
E^{syn}_{\rm \gamma, c,r}&\simeq& \frac{27\sqrt{2\pi}q_e\,m_e\xi^5}{8\sigma^2_T}\,(1+z)^{-3/2}\,(1+x_r+x_r^2)^{-2}\,\epsilon_{B,r}^{-3/2}\,A^{-2}\cr
&&\hspace{5.0cm}\times\,E^{1/2}\,T_{90}^{1/2}\,,
\eary
}
and the peak flux 
\be
F^{syn}_{\rm \gamma, max,r}\simeq\frac{\sqrt2 m_e\sigma_T}{64\,m_p\,q_e\xi^2}(1+z)^{2}\,\epsilon_{B,r}^{1/2}\,\Gamma^{-1}\,A^{1/2}\,D^{-2}\,E\,T_{90}^{-1}\,.
\ee
Synchrotron LCs are derived by \cite{2000ApJ...545..807K}.  Flux increases proportionally to $\sim t^{1/2}$ reaching a peak at $F^{syn}_{\rm \gamma,peak,r}\sim (E^{syn}_{\rm \gamma,r}/ E^{syn}_{\rm \gamma,c,r})^{-1/2}\,F^{syn}_{\rm \gamma, max,r}$ and after it starts decreasing as $\sim t^{-3}$ \citep{2003ApJ...597..455K}.
\paragraph{SSC emission}. Accelerated electrons can upscatter photons from low to high energies as
{\small
\bary\label{ic}
&&E^{\rm ssc}_{\rm \gamma,a,r}\sim2\gamma^2_{e,m,r}E^{syn}_{\rm \gamma, a,r},\hspace{1cm} E^{\rm ssc}_{\rm \gamma,m,r}\sim2\gamma^2_{e,m,r}E^{syn}_{\rm \gamma, m,r},\cr
&&E^{ssc}_{\rm \gamma,c,r}\sim2\gamma^2_{\rm e,c,r}\,E^{syn}_{\gamma,c,r},\hspace{0.3cm} {\rm and}\hspace{0.3cm}F^{ssc}_{\rm \gamma,max,r}\sim\,k\tau\,F^{syn}_{\rm \gamma,max,r}\,,
\eary
}
where {\small $k=4(p-1)/(p-2)$} and {\small $\tau=\frac{\sigma_T N(\gamma_e)}{4\pi r_d}$} is the optical depth of the shell. Here $N_e$ is the number of radiating electrons . From eqs. (\ref{ic}), (\ref{synrev}), (\ref{gamma_mr}) and (\ref{gamma_cr}), we get the break SSC energies     
{\small
\bary\label{sscrev}
E^{ssc}_{\rm \gamma,a,r} &\simeq& \frac{2^{61/5}\pi^{36/10} q_e^{8/5}\,m_p^{2/5}\,(p+2)^{3/5}\,(p-2)\,\xi^{9/5}}{5^{3/5}\,3^{14/5}\,\Gamma(5/6)^{3/5}\,(3p+2)^{3/5}(p-1)^{2/5}\,m_e^{2}}\cr
&&\hspace{0.4cm}\times\,(1+z)^{-19/10}\,\epsilon_{e,r}\,\epsilon_{B,r}^{1/5}\, E^{-19/10}\,A^{27/10}\,\Gamma^4 T^{9/10}_{90}\cr
E^{ssc}_{\rm \gamma, m,r}&\simeq& \frac{128\sqrt2\,\pi^{3/2}q_e\,m_p^4 (p-2)^4}{9\,m_e^5\,(p-1)^4}\,  (1+z)^{-1}\,\epsilon_{e,r}^4\,\epsilon_{B,r}^{1/2}\,\Gamma^{4}\cr
&&\hspace{5.5cm}\times\,A^{3/2}\,E^{-1}\cr
E^{ssc}_{\rm \gamma,c,r}&\simeq& \frac{0.85\,\,m_e^3\,q_e\xi^9}{\pi^{3/2}\,\sigma_T^4}  (1+z)^{-3/2}\,(1+x_r+x_r^2)^{-4}\,\epsilon_{B,r}^{-7/2}\cr
&&\hspace{4cm}\times\,\Gamma^{-6}\,A^{-6}\,E^{5/2}\,T_{90}^{1/2}\cr
F^{ssc}_{\rm \gamma,max,r}&\simeq&\frac{\sqrt2\sigma_T^2\,m_e\,(p-1)}{192\,m_p^2\,q_e\,\xi^4\,(p-2)}  (1+z)^{3}\,\epsilon_{B,r}^{1/2}\,\Gamma^{-2}\,A^{3/2}\cr
&&\hspace{4.5cm}\times\,D^{-2}\,E\,T_{90}^{-2}\,,
\eary
}
and the break energy at the KN regime is
\be\label{kn}
E^{KN}_{\gamma,r}=\frac{27m_e^2\,\xi^2}{64\pi\sigma_T}\,(1+z)^{-1}\,(1+x_r+x^2_r)^{-1}\,\epsilon_{B,r}^{-1}\,E\,\Gamma^{-2}\,A^{-2}\,.
\ee
At the shock crossing time $(t_d$), the SSC flux reaches the peak $F^{ssc}_{\rm \gamma,peak,r}\sim (E^{ssc}_{\rm \gamma,r}/ E^{ssc}_{\rm \gamma,c,r})^{-1/2}\,F^{ssc}_{\rm \gamma,max,r}$\citep{2003ApJ...597..455K}  at :
{\small
\bary\label{speak}
F^{ssc}_{\rm \gamma,peak,r}&\simeq&\frac{0.013\,m_e^3q_e(p-2)\xi^9}{\xi^{3/4}\,\sigma_T\,(p-1)} (1+z)^{-1/2}\,x_r\,(1+x_r+x^2_r)^{-5}\cr
&&\hspace{0.1cm}\times\, \epsilon_{e,r}\,\epsilon_{B,r}^{-7/2}\,\Gamma^{-6}\,A^{-6} D^{-2}\,E\,T_{90}^{-1/2}\,\left\{E^{ssc}_{\rm \gamma,r}\right\}^{-1/2}\,.
\eary
}
SSC LCs can be analytically derived from  \citet{2000ApJ...536..195C}.  For  $t < t_d$,  we take into account that: i) the number of radiating electrons N$_e$ and the spherical radius in the shocked shell region increase with time as N$_e\sim t$ and $r\sim t$, ii) the maximum flux of synchrotron is independent of time $F^{syn}_{\rm \gamma, max}\sim t^0$ and iii) the SSC cooling break energy  $E^{ssc}_{\rm \gamma, c}\sim \gamma^2_c  E^{syn}_{\rm \gamma, c}$ increases as $\sim t^3$, then the SSC flux increases as $F^{ssc}_{\rm \nu}\sim {E^{ssc}_{\rm \gamma, c}}^{1/2} F^{ssc}_{\rm \nu, max} \sim {E^{ssc}_{\rm \gamma, c}}^{1/2} \frac{N_e}{r_d^2} F^{syn}_{\rm \gamma, max}\sim t^{1/2}$.  For $t  > t_d$,  the SSC flux decreases as $F^{ssc}_{\rm \nu}\sim {E^{ssc}_{\rm \gamma, m}}^{-(p-1)/2} F^{ssc}_{\rm \nu, max} \sim {E^{ssc}_{\rm \gamma, m}}^{-\frac{p-1}{2}} \frac{N_e}{r_d^2} F^{syn}_{\rm \gamma, max}\sim t^{-\frac{p+1}{2}}$, where $E^{ssc}_{\rm \gamma, m}\sim \gamma^2_m  E^{syn}_{\rm \gamma, m}$ decreases as $\sim t^{-1}$.  It is valuable to note that the decay index of the emission for  $t  > t_d$ might be higher than $\frac{p-1}{2}$,   due to the angular time delay effect \citep{2003ApJ...597..455K}.
\subsubsection{Thin-shell case}
In the thin-shell case,  the RS cannot decelerate the shell effectively. The deceleration time and the minimum electron Lorentz factor are in the form
\be
 t_{dec}= \frac{9}{64\pi\,\xi^2}(1+z)\,E\,A^{-1}\,\Gamma^{-4}\,,
\ee
and  
\be\label{gamam_thin}
\gamma_{\rm e,m,r}=\epsilon_{e,r}\biggl(\frac{p-2}{p-1}\biggr) \frac {m_p} {m_e}\,,
\ee
respectively. Here the bulk Lorentz factor at the shock is $ \Gamma \simeq \Gamma_{d} <\Gamma_c$. 
\paragraph{Synchrotron emission}.  Performing a similar analysis of the thick-shell  case, from eqs. (\ref{synforw}), (\ref{param}),  (\ref{conec}) and (\ref{gamam_thin}),  we get that the synchrotron break energies and maximum flux can be written as  
{\small
\bary\label{synrev}
E^{syn}_{\rm \gamma,a,r} &\simeq& \frac{2^{34/5}\pi^{27/10}\, q_e^{8/5}\,(p+2)^{3/5}\,(p-1)^{8/5}}{5^{3/5}\,3\,\Gamma(5/6)^{3/5}\,(3p+2)^{3/5}(p-2)\,m_p^{8/5}}\cr
&&\hspace{1.3cm}\times\,(1+z)^{-1}\,\epsilon_{e,r}^{-1}\,\epsilon_{B,r}^{1/5}\, E^{-1}\,A^{9/5} \Gamma^{2/5}\cr
E^{syn}_{\rm \gamma,m,r}&\simeq& \frac{64\sqrt2\pi^{3/2}\,q_e\,m_p^2\,(p-2)^2}{9\,m_e^3\,(p-1)^2}\,(1+z)^{-1}\,\epsilon_{e,r}^{2}\,\epsilon_{B,r}^{1/2}\,\Gamma^{4}\,A^{3/2}\cr
&&\hspace{5.2cm}\times\,E^{-1} \cr
E^{syn}_{\rm \gamma,c,r}&\simeq& \frac{81\sqrt2  q_e\,m_e\,\xi^{4}}{64\,\sigma^2_T}\,(1+z)^{-1}\,(1+x_r+x_r^2)^{-2}\,\epsilon_{B,r}^{-3/2}\,E\cr
&&\hspace{3.8cm}\times\,A^{-5/2}\,\Gamma^{-2} \cr
F^{syn}_{\rm \gamma,max,r}&\simeq&\frac{\pi\sqrt2\,m_e\sigma_T}{9\,m_p\,q_e}(1+z)\,\epsilon_{B,r}^{1/2}\,D^{-2}\,\Gamma^3\,A^{3/2}\,.
\eary
}
\paragraph{SSC emission}.  In a like manner to the thick-shell case, from eqs. (\ref{ic}), (\ref{synrev}), (\ref{gamma_mr}) and (\ref{gamma_cr}),  we derive the break energies of SSC emission      
{\small
\bary\label{sscrev_thin}
E^{ssc}_{\rm \gamma,a,r} &\simeq& \frac{2^{34/5}\pi^{27/10}\, q_e^{8/5}\,m_p^{2/5}\,(p+2)^{3/5}\,(p-2)}{5^{3/5}\,3\,\Gamma(5/6)^{3/5}\,m_e^2\,(3p+2)^{3/5}(p-1)^{2/5}}\cr
&&\hspace{1.8cm}\times\,(1+z)^{-1}\,\epsilon_{e,r}\,\epsilon_{B,r}^{1/5}\, E^{-1}\,A^{9/5} \Gamma^{2/5}\cr
E^{ssc}_{\rm \gamma, m,r}&\simeq& \frac{128\sqrt2\pi^{3/2}\,q_e\,m_p^4\,(p-2)^4}{9\,m_e^5\,(p-1)^4}\,(1+z)^{-1}\,\epsilon_{e,r}^{4}\,\epsilon_{B,r}^{1/2}\,A^{3/2}\cr
&&\hspace{5.0cm}\times\,E^{-1}\,\Gamma^{4} \cr
E^{ssc}_{\rm \gamma, c,r}&\simeq& \frac{0.32\,  q_e\,m_e^3\,\xi^8}{\pi^2\,\sigma^4_T} (1+z)^{-1}\,(1+x_r+x_r^2)^{-4}\,\epsilon_{B,r}^{-7/2}\,E^3\cr
&&\hspace{4.5cm}\times\,A^{-13/2}\,\Gamma^{-8} \cr
F^{ssc}_{\rm \gamma,max,r}&\simeq&\frac{64\sqrt2\pi^2\,m_e\sigma^2_T\, (p-1)}{243\,q_e\,m^2_p(p-2)}  (1+z)\,\epsilon_{B,r}^{1/2}\,A^{7/2}\,D^{-2}\,\cr
&&\hspace{4.8cm}\times\,E^{-1}\, \Gamma^{6}\,.
\eary
}
The break energy at the KN regime is given by eq. (\ref{kn}). 
\section{Application: GRB 110731A}
GRB 110731A was detected on 2011 July 31 by both instruments aboard Fermi; Gamma-Ray Burst Monitor (GBM) and Large Area Telescope (LAT) \citep{2013ApJ...763...71A}, the three instruments aboard Swift; BAT, XRT and UVOT  \citep{2011GCNR..343....1O}  and ground-based observatories; Microlensing Observations in Astrophysics (MAO) telescope   \citep{2011GCN..12225...1T} and Gamma-ray Burst Optical/Near-Infrared Detector (GROND).\\
LAT localized GRB 110731A with coordinates R. A.=18$^h$41$^m$00$^s$ and dec.=-28$^\circ$31'00'' (J2000), with a 68\% confidence error radius of 0.2$^\circ$. Swift/BAT perceived immediately this burst after the detection  by both instruments of Fermi, whereas RXT and UVOT began observations 56s after the BAT trigger.  UVOT swiftly determined the afterglow position as R.A. = 18$^h$42$^m$00$^s$.99 and dec.=-28$^\circ$32'13''.8 (J2000), with a 90\% confidence. The lack of observation in the UV filters is consistent with the measured redshift z=2.83 \citep{2011GCN..12225...1T}. MAO observations started 3.3 minutes after the Swift trigger for GRB 110731A.  Using a 61 cm Boller \& Chivens telescope at the Mt. John University Observatory in New Zealand, I- and V-band images were collected 105 minutes after the trigger \citep{2011GCN..12242...1T}.  Finally, GROND  mounted on the 2.2 m MPG/ESO telescope at La Silla Observatory, Chile, imaged GRB 110731A  for 2.74 days after the trigger  \citep{2008PASP..120..405G}.\\
By considering the typical values of the stellar wind ($A=A_{\star}\times\,(5.0\times 10^{11})$ gram/cm with $A_{\star}=0.1$; \citet{2000ApJ...536..195C}, the parameter $\xi=0.56$; \citet{1998ApJ...493L..31P}) and those inferred by observations: redshift  $z=2.83$ \citep{2011GCN..12225...1T}, total energy $E\simeq10^{54}$ erg and duration of GRB $T_{90}$= 7.3 s, we will apply the leptonic model developed in this work to interpret the LAT  LC observations.\\ 
Taking into account the fact that the peak of the flux density was present at the end of the prompt phase; in the interval [5.47 s, 5.67 s], and that after the LAT flux decays smoothly during the whole temporally extended emission, we constrain the Lorentz factor $\Gamma\simeq 520 > \Gamma_c$ so that the deceleration time takes place at
\be
t_{dec}\simeq5.55\,{\rm s}\biggl(\frac{1+z}{4}\biggr)\,E_{54}\,A^{-1}_{\star,-1}\,\Gamma^{-4}_{2.72}\,,
\ee
and the RS evolves in the thick-shell case with  a critical Lorentz factor 
\bary\label{gamma_cr}
\Gamma_c&\simeq&472.5\biggl(\frac{1+z}{4}\biggr)^{1/4} \,A^{-1/4}_{\star,-1}\,E^{1/4}_{54}\left(\frac{T_{90}}{7.3s}\right)^{-1/4}\,.
\eary
\noindent We plot the synchrotron and SSC spectral breaks of the FS and RS as a function of equipartition parameters ($\epsilon_{B,f(r)}$ and $\epsilon_{e,f(r)}$), considering the typical values of the magnetic  ($10^{-5}\,\leq\epsilon_{B,f(r)}\leq\,1$) and electron ($\epsilon_{e,f(r)}$= 0.5, 0.1 and 0.05) parameters  \citep{2014ApJ...785...29S}, as shown in fig \ref{parameters}. We describe this figure as follows.\\
\paragraph{Synchrotron spectral breaks from FS (right-hand panel).} From this panel, we can see that the characteristic energy is an increasing  function of $\epsilon_{B,f}$ and $\epsilon_{e,f}$, the cooling energy is a decreasing function of $\epsilon_{B,f}$ and the self-absorption energy is an increasing  function of $\epsilon_{B,f}$ and a decreasing function of $\epsilon_{e,f}$.   The characteristic energy lies in the ranges:  {\small $5.3 \times 10^2\, {\rm eV} \leq E^{syn}_{\gamma, m,f}\leq  \,2.8\times 10^{5} \,{\rm eV}$}, {\small $3.5 \times 10^3\, {\rm eV} \leq E^{syn}_{\gamma, m,f}\leq  6.8 \times10^{5}\, {\rm eV}$ and  $6.1 \times 10^4\, {\rm eV} \leq E^{syn}_{\gamma,m,f}\leq  1.9 \times10^{7} \,{\rm eV}$} for $\epsilon_{e,f}$= 0.05, 0.1 and 0.5, respectively. Additionally, the cooling energy lies in the range  {\small $  8.9\times 10^{-6} \,{\rm eV} \leq	  E^{syn}_{\gamma, c,f}\leq 2.1 \times 10\, {\rm eV} $} and the synchrotron self-absorption energy in the ranges {\small $5.2 \times 10^{-3}\, {\rm eV} \leq E^{syn}_{\gamma, a,f}\leq  \,4.6\times 10^{-2} \,{\rm eV}$ and $5.1 \times 10^{-4}\, {\rm eV} \leq E^{syn}_{\gamma, a,f}\leq  \,6.3\times 10^{-3} \,{\rm eV}$}  for $\epsilon_{e,f}$= 0.05 and 0.5, respectively. Furthermore, one can see that self-absorption energy is in the weak self-absorption regime ($E^{syn}_{\gamma, a,f} < E^{syn}_{\gamma, c,f}$) for  $\epsilon_{B,f} <$ 0.003 (0.007) and $\epsilon_{e,f}$ = 0.05 (0.5), otherwise  the synchrotron spectrum would be in the strong absorption regime. \\
  \paragraph{Synchrotron spectral breaks from RS (left-hand panel).} From this panel, we can see that synchrotron spectral breaks from RS have the same behavior as FS although at lower energy ranges.  The characteristic synchrotron energy is in the ranges: {\small $0.01\,{\rm eV} \leq E^{syn}_{\gamma, m,r}\leq  5.1\, {\rm eV}$}, {\small $0.08\, {\rm eV} \leq E^{syn}_{\gamma, m,r}\leq   1.1 \times10^2 \,{\rm eV}$ and  $1.2\, {\rm eV} \leq E^{syn}_{\gamma, m,r}\leq  6.2 \times10^2 \,{\rm eV}$} for $\epsilon_{e,r}$=0.05, 0.1 and 0.5, respectively.  The cooling energy is in the range $1.8 \times 10^{-6}\, {\rm eV} \leq E^{syn}_{\gamma, c,r} <  46.2\, {\rm eV}$ and the self-absorption energy in the ranges $3.8 \times 10^{-7}\, {\rm eV} \leq E^{syn}_{\gamma, a,r}\leq  \,8.1\times 10^{-7} \,{\rm eV}$ and $7.3 \times 10^{-9}\, {\rm eV} \leq E^{syn}_{\gamma, a,r}\leq  \,6.8\times 10^{-8} \,{\rm eV}$  for $\epsilon_{e,r}$= 0.05 and 0.5, respectively.  It is important to say  that synchrotron self-absorption is in the weak absorption regime ($E^{syn}_{\gamma, a,r} < E^{syn}_{\gamma, c,r}$)  for any value of equipartition parameters considered here. However, as $E^{syn}_{\gamma, a,r}\propto A^{11/5}$  and $E^{syn}_{\gamma, c,r}\propto A^{-2}$ any significant increase of wind density, the absorption energy would become higher than cooling energy ($E^{syn}_{\gamma, a,f} > E^{syn}_{\gamma, c,f}$), then the synchrotron spectrum could change to the strong absorption regime. In this case, heating of low-energy electrons due to synchrotron absorption leads to pile-up of electrons, and there appears a thermal component besides the power-law spectrum \citep{2013MNRAS.435.2520G}.
\paragraph{SSC spectral breaks from RS (panel below).}    SSC spectral breaks have similar behavior as in the previous cases. The characteristic SSC energy lies in the range $52.3\, {\rm eV} \leq E^{ssc}_{\gamma, m,r}\leq  1.12 \times 10^{4}\, {\rm eV}$, $7.8 \times 10^3\, {\rm eV} \leq E^{ssc}_{\gamma, m,r}\leq  2.3 \times10^{5}\, {\rm eV}$ and  $6.2 \times 10^5\, {\rm eV} \leq E^{ssc}_{\gamma, m,r}\leq  1.17 \times10^{8}\, {\rm eV}$ for $\epsilon_{e,r}$=0.05, 0.1 and 0.5, respectively while the cooling energy lies in the range $8.2 \times 10^{-10}\, {\rm eV} \leq E^{ssc}_{\gamma, c,r} <  2.1\times10^{10}\, {\rm eV}$.\\
To obtain the values of parameters $\epsilon_{B,f(r)}$ and $\epsilon_{e,f(r)}$ that reproduce the multiwavelength LC observations, we use the method of Chi-square ($\chi^2$) minimization \citep{1997NIMPA.389...81B}.     We describe  LAT flux observations by synchrotron radiation from FS and SSC emission from RS; the whole temporally extended emission  using synchrotron LCs in the  fast-cooling regime (eq. \ref{fcsyn_t})  and the brightest peak at 5.5 s  with  SSC LCs  in the the fast-cooling regime  (eq. \ref{speak}),  for high-energy electrons radiating at $\simeq$ 100 MeV.  Following the analysis showed in  \cite{2013ApJ...763...71A},  we fit the X-ray ($t < 4.6\, {\rm ks}$) flux  with the synchrotron LC in the fast-cooling regime (eq. \ref{fcsyn_t}) at  t= 100 s for electrons radiating at  $E^{syn}_{\gamma,f}=$ 2 keV, and  optical and X-ray ($t > 4.6\, {\rm ks}$) fluxes with the LC in slow-cooling regime (eq. \ref{scsyn_t})  at t=600 s and 4000 s for $E^{syn}_{\gamma,f} =$ 10 eV and  0.7 keV, respectively.\\ 
We plot the values of parameters $\epsilon_{B,f(r)}$ and $\epsilon_{e,f(r)}$ for p=2.15, 2.2, 2.25 and 2.3 (see figs. \ref{fit_forward} and \ref{fit_peak}) that reproduce the multiwavelength LC observations (see fig. \ref{fit_afterglow}).  As shown in fig. \ref{fit_forward} the areas in yellow (green) colors show the set of parameters that describes the  temporally extended LAT (optical) flux and in blue (red) ones show those parameters that describe the X-ray for $t< 4.6\,{\rm ks}$ ($t > 4.6\, {\rm ks}$).  The areas in brown colors show the set of parameters that reproduce more than one flux at the same time.   For instance, as shown in the right-hand panel below (p=2.25), the equipartition parameters in the narrow strip between  0.38 $<\epsilon_{e,f}<$ 0.52 and  $10^{-4.5}\,<\epsilon_{B,f}\,<10^{-4}$ would reproduce the  temporally extended LAT, X-ray and optical fluxes.   From fig. \ref{fit_peak}, one can see the set of parameters that describe the brightest LAT peak. Also it can be seen that as the electron spectral index increases the set of parameters is shifted to the right.\\ 
Given the values $\epsilon_{B,f}=10^{-4.15}$ and $\epsilon_{e,f}=0.40$ for p=2.25,  from eqs. (\ref{synforw}) and (\ref{t0}) we get that the synchrotron spectral breaks from FS are 
{\small
\bary\label{synforw_a}
E^{syn}_{\rm \gamma,a,f} &\simeq&  5.56\times 10^{-4}\,{\rm eV}\, \biggl(\frac{1+z}{4}\biggr)^{-2/5}\,\epsilon_{e,f,-0.4}^{-1}\,\epsilon_{B,f,-4.15}^{1/5}\cr
&&\hspace{4.3cm}\times\,A^{6/5}_{\star,-1}\,E^{-2/5}_{54}\,  t_{1}^{-3/5}\cr
E^{syn}_{\rm \gamma,m,f} &\simeq&  77.45\,{\rm keV}\, \biggl(\frac{1+z}{4}\biggr)^{1/2}\,\epsilon_{e,f,-0.4}^2\,\epsilon_{B,f,-4.15}^{1/2}\,E^{1/2}_{54}\,  t_{1}^{-3/2}\cr
E^{syn}_{\rm \gamma,c,f}  &\simeq&  0.30\, {\rm eV}\, \biggl(\frac{1+z}{4}\biggr)^{-3/2}\,\biggl(\frac{1+x_f}{11}\biggr)^{-2}\, \epsilon_{B,f,-4.15}^{-3/2}\,A^{-2}_{\star,-1}\cr
&&\hspace{4.6cm}\times\, E^{1/2}_{54}\, t_{1}^{1/2}\, \cr
E^{syn}_{\rm \gamma,max,f}  &\simeq&36.94\, {\rm GeV} \,\biggl(\frac{1+z}{4}\biggr)^{-3/4}\,E^{1/4}_{54}\,A^{-1/4}_{\star,-1}\,t^{-1/4}_1\,,
\eary
}
\noindent and the transition time from fast- to slow-cooling regime is
\be
t^{syn}_0=123.46 s \,\biggl(\frac{1+z}{4}\biggr)\epsilon_{e,f,-0.4}\,\epsilon_{B,f,-4.15}\,A_{\star,-1}\,.
\ee
It is important to highlight that the maximum photon energy achieved by synchrotron radiation is $E^{syn}_{\rm \gamma,max,f} \simeq 36.94\, {\rm GeV}$ for t=10s and $E^{syn}_{\rm \gamma,max,f} \simeq 20.77\, {\rm GeV}$ for t=100s.\\
From eq. (\ref{scic}) we get that the SSC scattering break energies are
{\small
\bary\label{sscforw_a}
E^{ssc}_{\rm \gamma, m,f} &\simeq&  11.66\,{\rm TeV}\biggl(\frac{1+z}{4}\biggr)\,\epsilon_{e,f,-0.4}^4\,\epsilon_{B,f,-4.15}^{1/2}\cr
&&\hspace{4.3cm}\times\,A^{-1/2}_{\star,-1}\,E_{54}\,  t_{2}^{-2}\cr
E^{ssc}_{\rm \gamma, c,f}  &\simeq&  162.8\, {\rm keV}\, \biggl(\frac{1+z}{4}\biggr)^{-3}\,\biggl(\frac{1+x_f}{11}\biggr)^{-4}\,\epsilon_{B,f,-4.15}^{-7/2}\cr
&&\hspace{4.7cm}\times\, A^{-9/2}_{\star,-1}\,E_{54}\, t_{2}^{2}\, \cr
F^{ssc}_{\rm \gamma, max,f}&\simeq&  0.41\,{\rm Jy}\biggl(\frac{1+z}{4}\biggr)\,\epsilon_{B,f,-4.15}^{1/2} \,A^{1/2}_{\star,-1}\,D^{-2}_{28}E_{54}\,.\hspace{0.2cm}
\eary
}
The break energy at the KN regime is $E^{KN}_{\gamma,f}= 42.33\, {\rm GeV}$. From the SSC LCs (eqs. \ref{fcic_t} and \ref{scic_t}), one can see that although the temporal power index of LC for $E^{ssc}_{\rm \gamma,m,f}<E^{ssc}_\gamma < E^{ssc}_{\rm \gamma,max,f}$ would be in agreement with the temporally extended  LAT flux, the energy range would not.  The detection of the photon with energy 3.4 GeV at $\sim$ 436 s could be explained by synchrotron radiation as well as Compton scattering emission.
Considering  $\epsilon_{e,r}=\epsilon_{e,f}= 0.4$, we obtain $\epsilon_{B,f}= 0.28$ for p=2.25.  From eqs. (\ref{synrev}) and (\ref{sscrev}) we get that the synchrotron and SSC break energies are
{\small
\bary\label{synrev_a}
E^{syn}_{\rm \gamma,a,r}&\simeq& 4.28\times 10^{-8} \, {\rm eV}\, \biggl(\frac{1+z}{4}\biggr)^{-7/5}\,\epsilon_{e,r,-0.4}^{-1}\,\epsilon_{B,r,-0.55}^{1/5}\,\Gamma^{2}_{2.72}\cr
&&\hspace{3.0cm}\times A_{\star,-1}^{11/5}\,E^{-7/5}_{54}\,\left(\frac{T_{90}}{7.3s}\right)^{2/5} \cr
E^{syn}_{\rm \gamma,m,r}&\simeq& 128.94  \, {\rm eV}\, \biggl(\frac{1+z}{4}\biggr)^{-1/2}\,\epsilon_{e,r,-0.4}^{2}\,\epsilon_{B,r,-0.55}^{1/2}\,\Gamma^{2}_{2.72}\cr
&&\hspace{3.0cm}\times A_{\star,-1}\,E^{-1/2}_{54}\,\left(\frac{T_{90}}{7.3s}\right)^{-1/2} \cr
E^{syn}_{\rm \gamma,c,r}&\simeq& 0.93\times 10^{-5} \, {\rm eV}\,  \biggl(\frac{1+z}{4}\biggr)^{-3/2}\,\biggl(\frac{1+x_r+x_r^2}{3}\biggr)^{-2}\cr
&&\hspace{2.2cm}\times\, \epsilon_{B,r,-0.55}^{-3/2}\,A^{-2}_{\star,-1}\,E^{1/2}_{54}\,\left(\frac{T_{90}}{7.3s}\right)^{1/2} \cr
F_{\rm \gamma,max,r}&\simeq&  4.31\times10^4  \,{\rm \,Jy}\,  \biggl(\frac{1+z}{4}\biggr)^{2}\,\epsilon_{B,r,-0.55}^{1/2}\,\Gamma^{-1}_{2.72}\,A^{1/2}_{\star,-1}\cr
&&\hspace{3.0cm}\times \,D^{-2}_{28}\,E_{54}\,\left(\frac{T_{90}}{7.3s}\right)^{-1}\,,
\eary
}
and 
{\small
\bary\label{ssc_a}
E^{ssc}_{\rm \gamma,m,r}&\simeq& 103.55 \, {\rm \ MeV}\,  \biggl(\frac{1+z}{4}\biggr)^{-1}\,\epsilon_{e,r,-0.4}^4\,\epsilon_{B,r,-0.55}^{1/2}\,\Gamma^{4}_{2.72}\cr
&&\hspace{4.8cm}\times  A^{3/2}_{\star,-1}\,E^{-1}_{54},\cr
E^{ssc}_{\rm \gamma,c,r}&\simeq&  5.86\times 10^{-3} \, {\rm \ eV}\, \biggl(\frac{1+z}{4}\biggr)^{-3/2}\,\biggl(\frac{1+x_r+x_r^2}{3}\biggr)^{-4}\cr
&&\hspace{1.6cm}\times \epsilon_{B,r,-0.55}^{-7/2}\,\Gamma^{-6}_{2.72}\,A^{-6}_{\star,-1}\,E^{5/2}_{54}\,\left(\frac{T_{90}}{7.3s}\right)^{1/2},\cr
F^{ssc}_{\rm \gamma,max,r}&\simeq& 1.42\times10^2 \,{\rm \,Jy}\, \biggl(\frac{1+z}{4}\biggr)^{3}\,\epsilon_{B,r,-0.55}^{1/2}\,\Gamma^{-2}_{2.72}\,A^{3/2}_{\star,-1}\cr
&&\hspace{3.2cm}\times D^{-2}_{28}\,E_{54}\,\left(\frac{T_{90}}{7.3s}\right)^{-2}\,,
\eary
}
respectively.  The break energy at the KN regime is $E^{KN}_{\gamma,r}= 52.71\, {\rm GeV}$.
\section{Conclusions}
\begin{figure}
\vspace{0.5cm}
{\centering
\resizebox*{1.1\textwidth}{0.7\textheight}
{\includegraphics{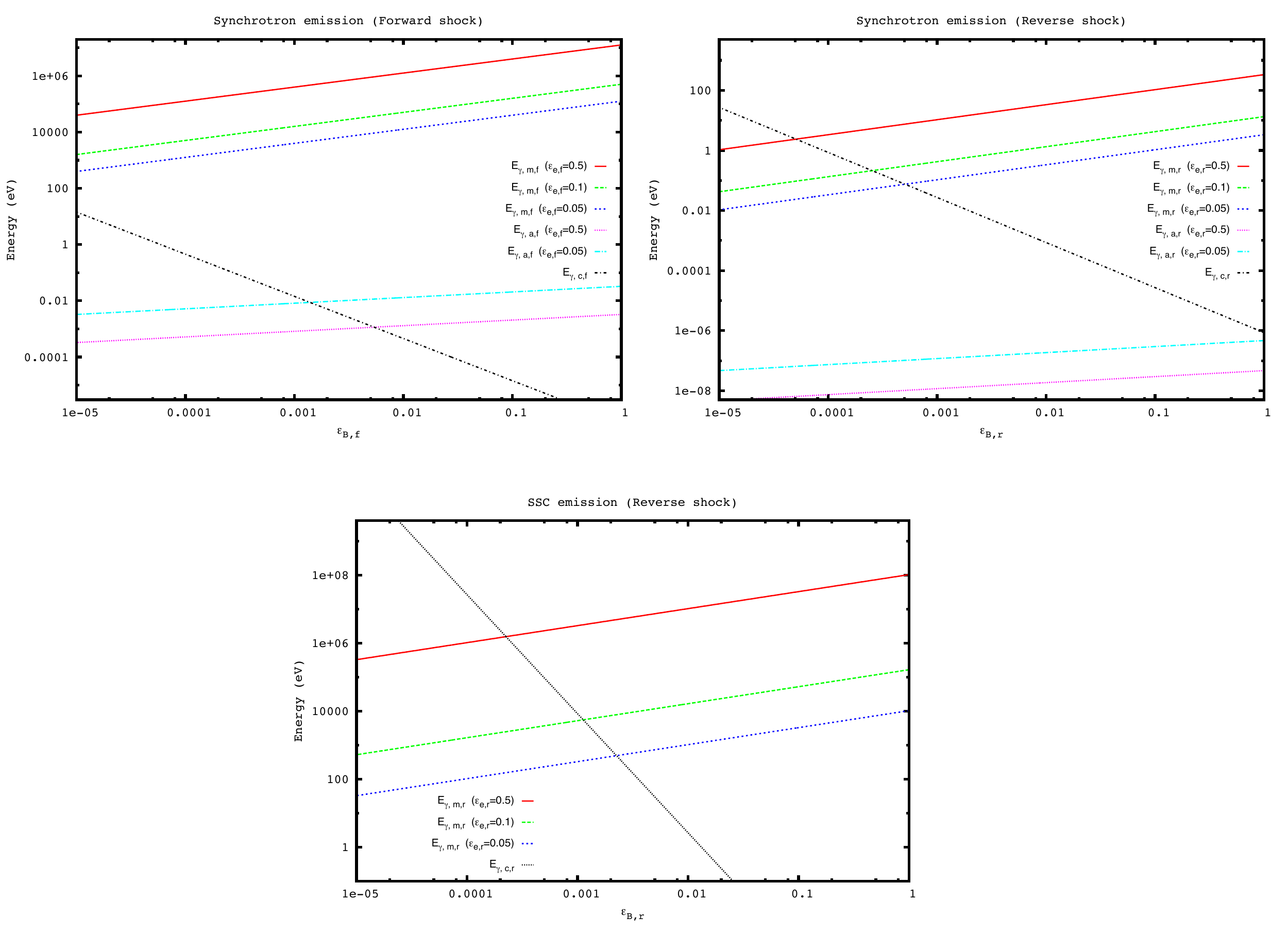}}
}
\caption{Break photon energies of synchrotron radiation from the forward (left-hand figure above) and reverse (right-hand figure above) shocks,  and  SSC emission from the RS (figure below)  as a function of magnetic equipartition parameter ($\epsilon_{B,f/r}$) for $\epsilon_{e,f/r}$ = 0.5, 0.1, 0.05 and 0.01.}
\label{parameters}
\end{figure}

\begin{figure}
\vspace{0.5cm}
{\centering
\resizebox*{1.05\textwidth}{0.6\textheight}
{\includegraphics{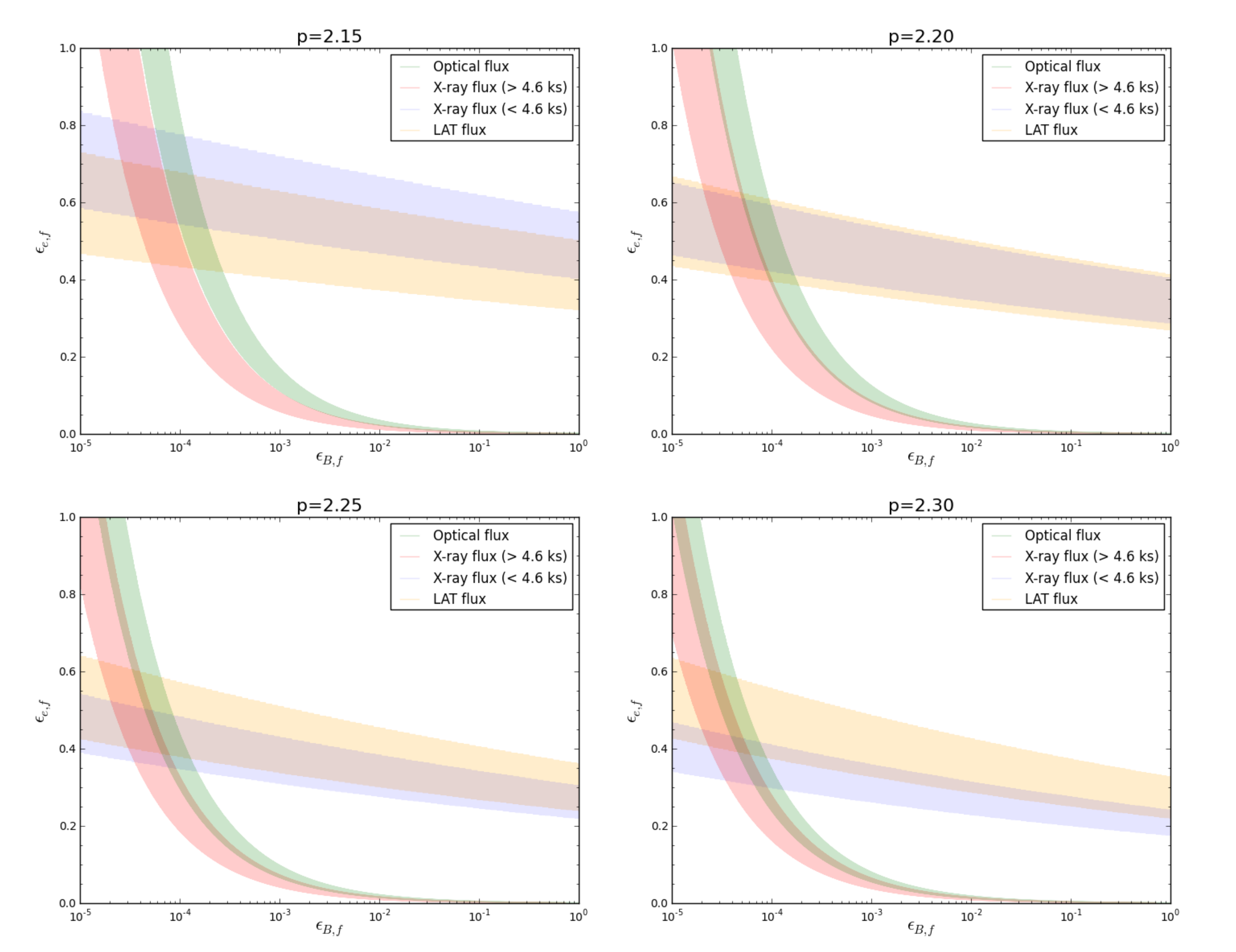}}
}
\caption{Values of equipartition parameters ($\epsilon_{B,f}$ and $\epsilon_{e,f}$) that reproduce the temporally extended LAT, X-ray and optical emission  through synchrotron radiation from FS.}
\label{fit_forward}
\end{figure}

\begin{figure}
\vspace{0.5cm}
{\centering
\resizebox*{1.05\textwidth}{0.6\textheight}
{\includegraphics{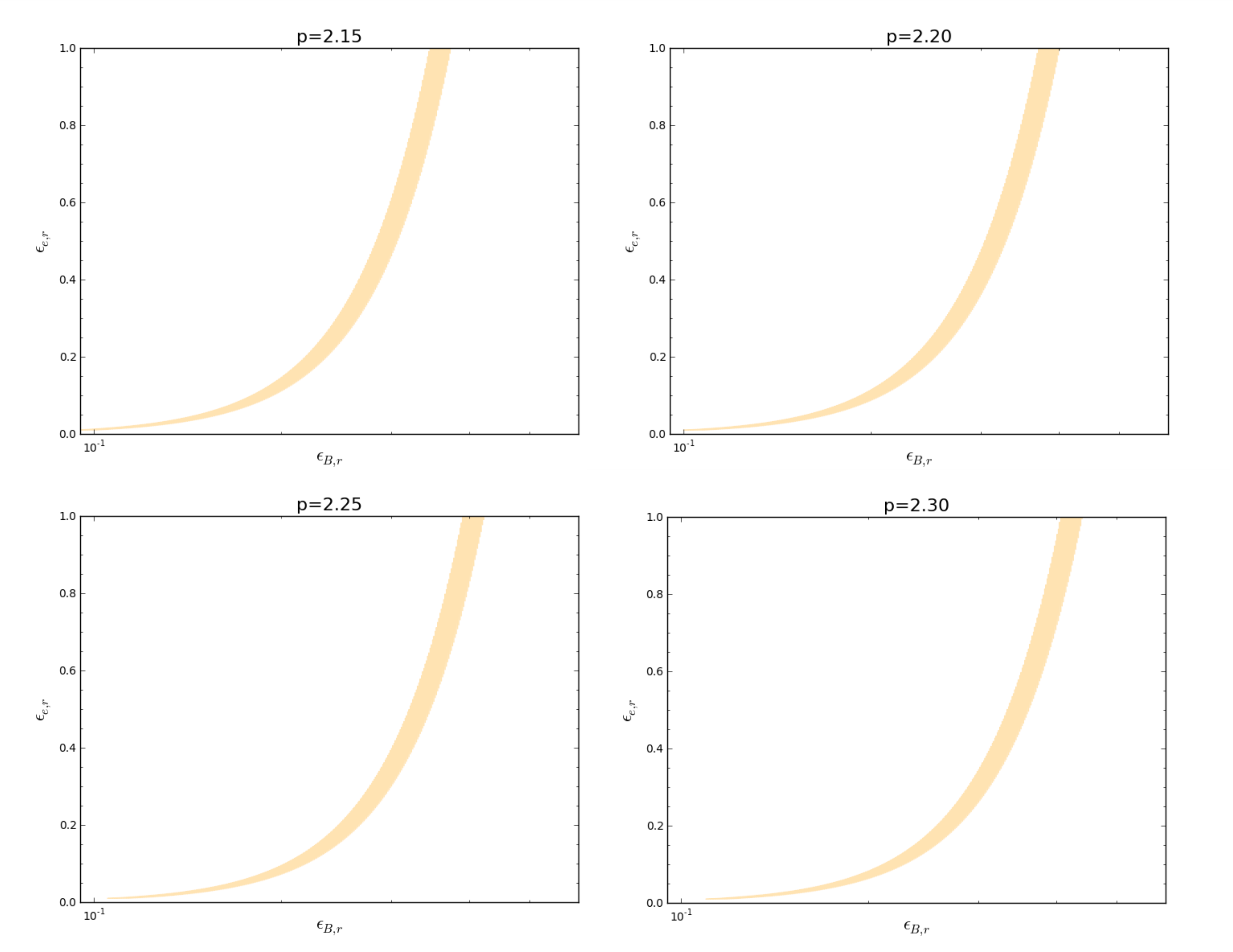}}
}
\caption{Values of equipartition parameters ($\epsilon_{B,r}$ and $\epsilon_{e,r}$) that describe the brightest  LAT peak through SSC emission from RS.}
\label{fit_peak}
\end{figure}

\begin{figure}
\vspace{0.5cm}
{\centering
\resizebox*{1\textwidth}{0.5\textheight}
{\includegraphics{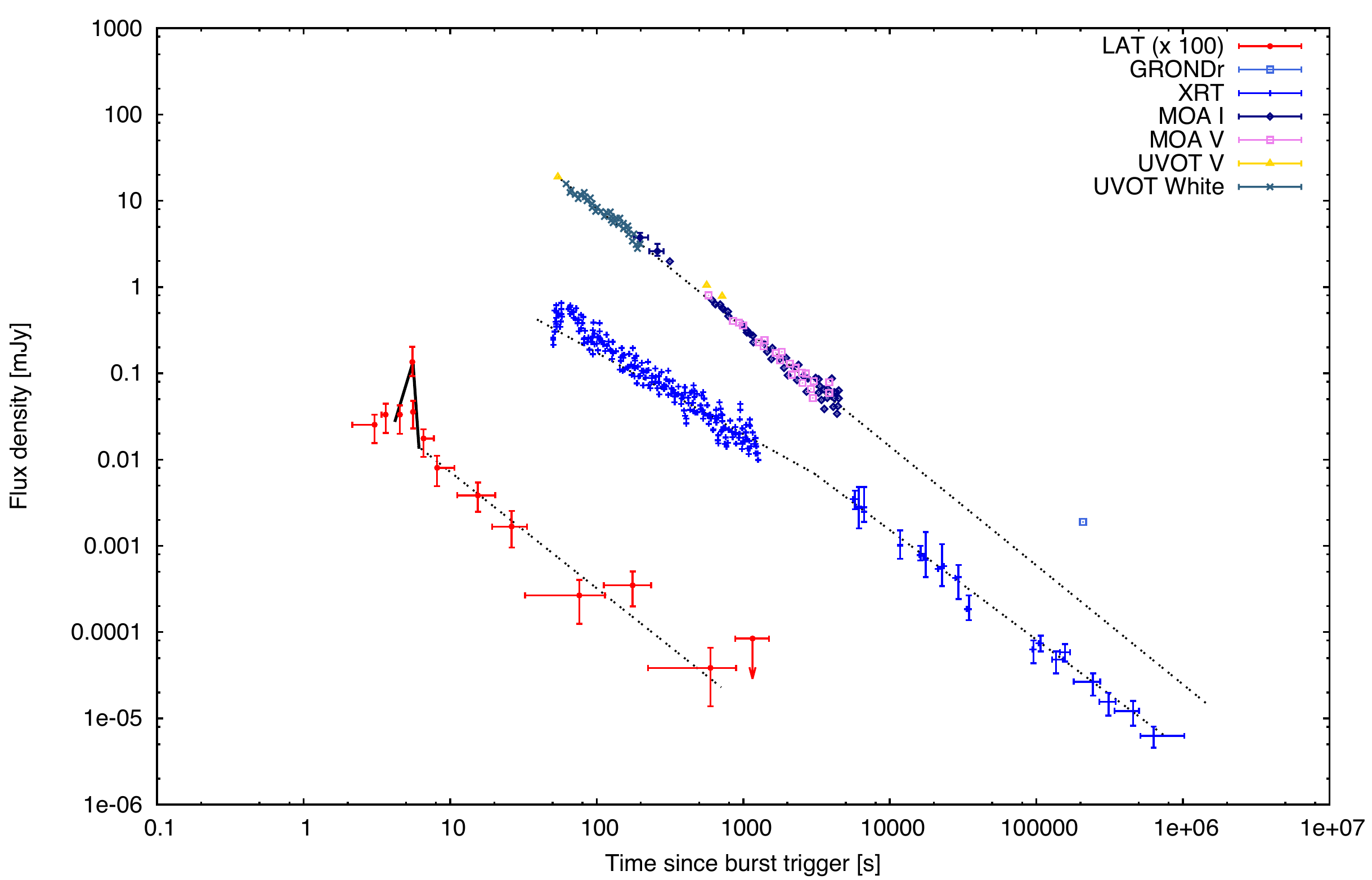}}
}
\caption{Fits of the multiwavelength LCs of GRB 110731A observation with our model. We use the RS  in the thick-shell regime to describe the brightest peak of the LAT  flux (continuos line) and the FS to explain  the temporally extended LAT, X-ray and optical emissions (dashed lines).}
\label{fit_afterglow}
\end{figure}

We have presented a leptonic model based on the evolution of an early afterglow in the stellar wind. We apply this model to describe the temporally extended emission and the brightest peak present in the LAT light curve of GRB 110731A, although additionally we fit the X-ray and optical light curve afterglows. \\ 
We consider that the ejecta propagating in the stellar wind is early decelerated at $\sim 5.5$ s and the RS evolves in the thick shell regime.   Taking into account  the values of redshift  $z=2.83$ \citep{2011GCN..12225...1T}, energy $E\simeq10^{54}$ erg, duration of GRB $T_{90}$= 7.3 s  \citep{2013ApJ...763...71A,2011GCNR..343....1O, 2011GCN..12225...1T, 2011GCN..12242...1T} and the stellar wind $A=5.0\times 10^{10}$ gram/cm \citep{2000ApJ...536..195C}, we get that the value of the Lorentz factor is $\Gamma\simeq 520$.\\
We plot the SSC and synchrotron spectral breaks from FS and RS as a function of equipartition parameters, as shown in fig \ref{parameters}.  We found that the synchrotron  self-absorption energies from FS  and RS are in the weak self-absorption regime  for  $\epsilon_{B,f} <$ 0.003 (0.007) and $\epsilon_{e,f}$ = 0.05 (0.5) and for  $ 10^{-5}\,< \epsilon_{B,r} < 1$ and $0.05\,< \epsilon_{e,r}\, < 0.5$ (considered here), respectively.  It is important to say that for {\small $\gamma_{e,a}\gg \gamma_{e,m}$} (not considered in this work) the synchrotron spectrum would be in the strong absorption regime ({\small $E^{syn}_{\rm \gamma,a,f}>E^{syn}_{\rm \gamma,c,f}$}). In this case a thermal peak due to  pile-up of electrons would appear around  {\small $E^{syn}_{\rm \gamma,a,f}\sim 6.23\, {\rm eV}\,\epsilon_{e,r,-0.4}^{2/7}\,\epsilon_{B,r,-0.55}^{1/14}\,A^{2/7}_{\star,-1}\,E^{1/14}_{54}\,\left(\frac{T_{90}}{7.3s}\right)^{-11/14}$} in the synchrotron spectrum of the RS, modifying the broken power-law spectrum \citep{2004ApJ...601L..13K,2013MNRAS.435.2520G}.\\
To find the equipartition parameters $\epsilon_{B,f(r)}$ and $\epsilon_{e,f(r)}$  we fit the multiwavelength afterglow LCs  (see fig. \ref{fit_afterglow}); the brightest LAT peak by SSC emission from RS and the extended temporally emissions (LAT, X-ray and optical) by synchrotron radiation from the FS. We  plot the set of values $\epsilon_{B,f(r)}$ and $\epsilon_{e,f(r)}$ that describes these observations, as shown in figs.  \ref{fit_forward} and  \ref{fit_peak}. From the FS, we chose the values  $\epsilon_{B,f}= 10^{-4.15}$ and $\epsilon_{e,f}=0.4$ for p=2.25, then we get the synchrotron spectral breaks: $E^{syn}_{\rm \gamma,a,f} \simeq  5.56\times 10^{-4}\,{\rm eV}$, $E^{syn}_{\rm \gamma,m,f} \simeq  77.45\,{\rm keV}$, $E^{syn}_{\rm \gamma,c,f}  \simeq  0.30\, {\rm eV}$ and $E^{syn}_{\rm \gamma,max,f}  \simeq 36.94\, {\rm GeV}$, and SSC spectral breaks: $E^{ssc}_{\rm \gamma, m,f} \simeq  11.66\,{\rm TeV}$ and $E^{ssc}_{\rm \gamma, c,f}  \simeq  162.8\, {\rm keV}$.  From the RS, we consider the values  $\epsilon_{e,r}=\epsilon_{e,f}= 0.4$ and $\epsilon_{B,r}= 0.28$, then  we obtain the synchrotron spectral breaks: $E^{syn}_{\rm \gamma,a,r}\simeq 4.28\times 10^{-8} \, {\rm eV}$, $E^{syn}_{\rm \gamma,m,r}\simeq 128.94  \, {\rm eV}$ and $E^{syn}_{\rm \gamma,c,r}\simeq 0.93\times 10^{-5}\,  {\rm eV}$, and SSC spectral breaks: $E^{ssc}_{\rm \gamma,m,r}\simeq 103.55 \, {\rm MeV}$ and $E^{ssc}_{\rm \gamma,c,r}\simeq  5.86\times 10^{-3} \, {\rm eV}$.\\
The current model accounts for the main temporal and spectral characteristics of GRB 110731A, having six free parameters (bulk Lorentz factor, density of the stellar wind, electron and magnetic equipartition parameters). From the parameters required and observed ($\Gamma$, $A$, $E$, $T_{90}$, $p$ and $z$), and the values obtained after describing the afterglow LC  ($\epsilon_{e,f}=\epsilon_{e,r}$ and  $\epsilon_{B,f}\neq \epsilon_{B,r}$), we can see that the difference between the SSC and synchrotron spectral breaks achieved in forward and RSs can be explained through the energy distribution given to magnetic field.  Comparing the magnetic equipartition parameters that best describe the emission at forward and RSs, we can see that magnetic fields in both shocks are related by  $B_f\simeq2\times 10^{-2}\,B_r$. The previous result as found in GRB 990123, GRB 021211, GRB 980923 and GRB 090926A  \citep{2003ApJ...595..950Z,2012ApJ...751...33F,2012ApJ...755..127S} illustrates that the magnetic field in the reverse-shock region is stronger ($\sim$ 50 times) than in the forward-shock region which indicates that the ejecta is magnetized. Unlike the FS emission that continue later at lower energies, the RS is shown as a single peak.  After the RS has crossed the ejecta, there are no freshly accelerated electrons injected and the emission drops sharply. The magnetization of the blast wave modifies the temporal characteristics of the brightest LAT peak; it becomes  much shorter in duration, for GRB 110731A the peak duration  is $\simeq 1s$  \citep{2004A&A...424..477F}.\\   
Some authors have claimed that the GeV emission detected by LAT during the prompt phase (before T$_{90}$) has an internal origin similar to its MeV counterpart \citep{2011MNRAS.415...77M, 2011ApJ...730..141Z, 2011ApJ...733...22H,  2011ApJ...730....1L}. However, in particular for GRB 110731A, we describe the multiwavelength data with synchrotron radiation as originating from FS in stellar medium starting at $\sim$ 5.5 s. Then, it is natural to think that the brightest LAT peak around the afterglow onset time comes from the RS as was explained in this work.  We note that bursts with sub-TeV photons at hundreds of seconds from the prompt phase could be  described by SSC emission from FS and be candidates to be detected by TeV $\gamma$-ray observatories as the High Altitude Water Cherenkov observatory (HAWC) \citep{2012APh....35..641A,2014arXiv1410.1536A}.\\
In summary,  we model not only  the LAT light curve (the long-lasting GeV emission extended up to 853 s  by synchrotron emission from FS and the brightest peak by SSC emission from RS) but also explain the multiwavelength afterglow observations in GRB 110731A using the leptonic model based on an early afterglow which evolves in a stellar wind.  The bulk Lorentz factor required in this model is $\Gamma\simeq520$ and the ejecta must be magnetized.  In this model the onset of the HE emission is delayed because it is emitted from external shocks.\\
\acknowledgments
We thank Bing Zhang, William Lee, Bin-Bin Zhang, Magdalena Gonz\'alez, Rodrigo Sacahui, Sylvain Guiriec and Peter Veres for useful discussions. This work was supported by Luc Binette scholarship and the projects IG100414 and Conacyt 101958.
%
%

%
\clearpage
\clearpage
\end{document}